\documentclass{article}
\usepackage{ijcai25}

\usepackage{times}
\usepackage{soul}
\usepackage{url}
\usepackage[hidelinks]{hyperref}
\usepackage[utf8]{inputenc}
\usepackage[small]{caption}
\usepackage{bm}
\usepackage{graphicx}
\usepackage{amsmath}
\usepackage{amsthm}
\usepackage{booktabs}
\usepackage{algorithm}
\usepackage{algorithmic}
\usepackage[switch]{lineno}
\usepackage{enumitem}
\usepackage{multirow}

\usepackage{amssymb}

\usepackage{caption}   %
\usepackage{color,xspace}

\usepackage{xcolor} 

\usepackage{multirow}  
\newcommand{\name}{\textsc{FMRec}\xspace}
\usepackage{tabularx}
\usepackage{tikz}
\usepackage{adjustbox}

\title{Flow Matching based Sequential Recommender Model}

\author{
Feng Liu$^1$
\and
Lixin Zou$^1$\thanks{Corresponding Author.} \and
Xiangyu Zhao$^2$ \and
Min Tang$^3$\and
Liming Dong$^4$\and \\
Dan Luo$^5$\and
Xiangyang Luo$^{6\ast}$\And
Chenliang Li$^1$ \\
\affiliations
$^1$Key Laboratory of Aerospace Information Security and Trusted Computing, Ministry of Education, School of Cyber Science and Engineering, Wuhan University \\
$^2$City University of Hong Kong, $^3$Monash University, $^4$National Defense University, $^5$Lehigh University \\
$^6$State Key Lab of Mathematical Engineering and Advanced Computing\\
\emails
\{liufeng.tanh, zoulixin\}@whu.edu.cn,
xianzhao@cityu.edu.hk,
min.tang@monash.edu,
dlm14@tsinghua.org.cn,
danluo.ir@gmail.com,
xiangyangluo@126.com,
cllee@whu.edu.cn
}

\begin{document}

\maketitle

\begin{abstract}
Generative models, particularly diffusion model, have emerged as powerful tools for sequential recommendation. 
However, accurately modeling user preferences remains challenging due to the noise perturbations inherent in the forward and reverse processes of diffusion-based methods. 
Towards this end, this study introduces \name, a Flow matching based model that employs a straight flow trajectory and a modified loss tailored for the recommendation task.
Additionally, from the diffusion-model perspective, we integrate a reconstruction loss to improve robustness against noise perturbations, thereby retaining user preferences during the forward process. 
In the reverse process, we employ a \textit{deterministic reverse sampler}, specifically an ODE-based updating function, to eliminate unnecessary randomness, thereby ensuring that the generated recommendations closely align with user needs.
Extensive evaluations on four benchmark datasets reveal that \name achieves an average improvement of \textbf{6.53\%} over state-of-the-art methods. 
The replication code is available at \url{https://github.com/FengLiu-1/FMRec}.

\end{abstract}

\section{Introduction}
\label{sec:introduction}

Diffusion model (DM), owned to their great ability to generate high-quality image~\cite{nichol2021improved,song2020denoising}, video \cite{ho2022video,harvey2022flexible,yu2024fastervd} and text \cite{gong2022diffuseq,wu2023ar}, has inspired the development of innovative adaptations in sequential recommendation systems~(\textit{e.g.}, DiffuRec~\cite{li2023diffurec} and DiffRec~\cite{wang2023diffusion}).
Usually, the diffusion-based model consists of two main phases: the forward procedure and the reverse procedure.
During the forward procedure of diffusion model, \textit{i.e.}, the \textbf{training procedure}, the model progressively adds noise to the real data based on a predefined \textit{noise schedule}, eventually transforming it into random noise resembling that drawn from a normal distribution. 
In contrast, the \textbf{reverse procedure}, or the inference stage, iteratively removes the noise from the sampled noise using a \textit{reverse samplers}, \textit{i.e.}, the SDE-based stochastic reverse sampler~\cite{nakkiran2024step}.
This process is typically conditioned on both the random noise and additional inputs, allowing for the generation of realistic samples.

\begin{figure}[!tbp]
\centering\includegraphics[width=0.45\textwidth]{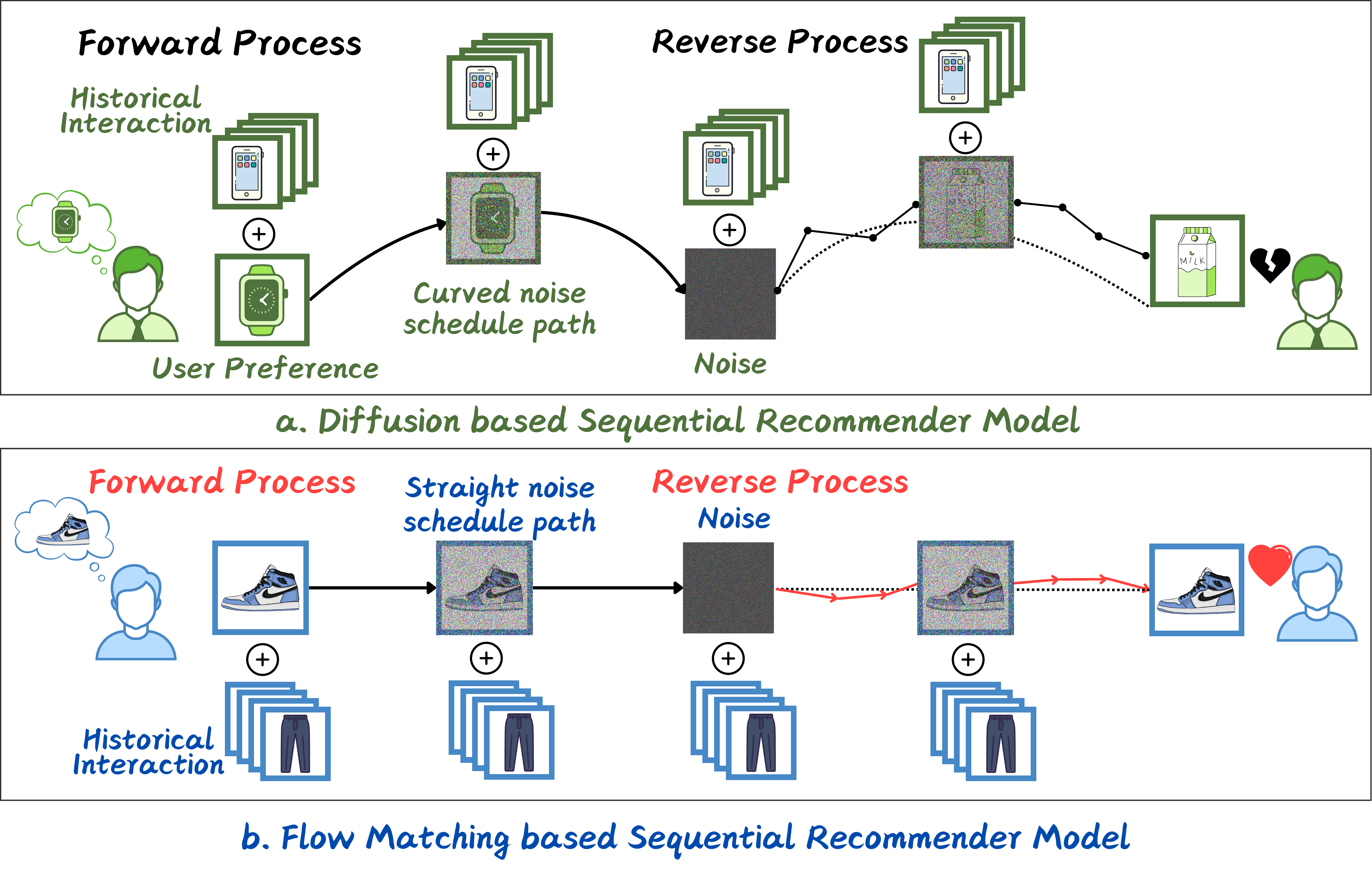}
\vspace{-0.2cm}
    \caption{An illustration that highlights the differences between Diffusion-based (a) and Flow Matching based (b) sequential recommender models in both the forward and reverse processes.}\label{fig:flowmatching_introduction}
    \vspace{-0.4cm}
\end{figure}

Following this paradigm, methods such as DiffuRec~\cite{li2023diffurec}, DreamRec~\cite{yang2024generate}, and DimeRec~\cite{li2024dimerec} have extended the diffusion model to the sequential recommendation. Specifically, these approaches generate next-item predictions by leveraging both random noise and user-item interactions. In the forward procedure, illustrated in Figure~\ref{fig:flowmatching_introduction}(a), these models progressively add noise to the target recommendation, transforming the actual next item into random noise. 
After integrating the random noise and historical interactions using a deep learning model, the reverse process utilizes 
a \textit{stochastic reverse sampler} to progressively denoise the next item from the noise-perturbed user preferences.

Although effective, existing methods exhibit several limitations:
\textbf{(1) Inaccurate user preference modeling}: 
The integration of random noise during the forward process, along with user preferences, can indeed compromise the accuracy of user preference modeling.
Additionally, these methods~\cite{li2023diffurec,wang2023diffusion} typically utilize a \textit{curved noise schedule path} during the noise addition process, which can result in error accumulation due to their long-curved trajectory, as illustrated by the curve in Figure~\ref{fig:flowmatching_introduction}(a). 
Consequently, during the reverse inference process, existing methods need to fit these curved paths, necessitating a greater number of diffusion steps to counteract the impact of these errors effectively.
While existing methods might be aware of this issue, they often address it by selecting hyperparameters that introduce minimal random noise into user preference modeling. 
However, this merely treats the symptoms rather than resolving the underlying problem.
\textbf{(2) Randomness on recommendation generation}: 
The \textit{stochastic reverse sampler} used in reverse procedures introduces randomness into the recommendation generation process, potentially resulting in irrelevant suggestions. 
These stochastic samplers typically introduce extra noise-perturbations during the sampling phase, yielding diverse and varied samples that are beneficial in tasks like image generation (\textit{e.g.}, creating different cat breeds such as Ragdolls, Persians, and Folded-ear cats).  
However, in sequential recommendation systems, the primary goal is to accurately predict the next likely item while exploring diverse yet relevant topics. 
Unfortunately, these unintended perturbations often lead to irrelevant recommendations, which can ultimately degrade the user experience. 
In the example shown in Figure~\ref{fig:flowmatching_introduction}(a), such perturbations can shift recommendations from ``watch'' to ``milk'' and negatively impact users' preferences on the actual platform.

Towards this end, this work firstly adapts Flow Fatching, \textit{i.e.}, a simplified diffusion model, for the sequential recommendation and proposes \name.
Specifically, in the forward process, we utilize a straight flow trajectory and derive a noise-free equivalent learning target for the sequential recommendation~(the theoretical analysis of straight trajectory's advantage is provided in Appendix~\ref{sec:Theoretical_appendix}).
Thereby, \name minimizes error accumulation and achieves more precise recommendations.
Additionally, we introduce a decoder architecture to reconstruct users' historical preferences, along with a corresponding \textit{interaction information reconstruction loss} during the training process to enhance the model's robustness against noise perturbations.
To control randomness in item generation, we implement a \textit{deterministic reverse sampler} using an ODE-based deterministic sampling method, effectively eliminating random perturbations during inference.
Figure~\ref{fig:flowmatching_introduction}(b) illustrates our proposed method. 
Finally, we conduct extensive experiments on four benchmark datasets and compare \name with state-of-the-art~(SOTA) approaches. 
An average improvement of \textbf{6.53\%} over SOTA verifies the effectiveness of our proposed methods. 

\section{Related Work}
\label{sec:related_work}
\subsection{Sequential Recommendation Systems}

The rapid advancement of deep learning has significantly enhanced sequential recommendation systems~\cite{zou2019reinforcement,tang2024unbiased} through various architectures. Early pioneering works employ Recurrent Neural Network (RNN)~\cite{donkers2017sequential,devooght2017long}, such as GRU4Rec~\cite{hidasi2015session}, RRN~\cite{wu2017recurrent}, and Convolutional Neural Networks (CNN)~\cite{de2018news}, including RCNN~\cite{xu2019recurrent} and Caser~\cite{tang2018personalized}, which effectively capture user preferences and immediate interests. 
More recent approaches have enhanced sequence modeling using self-attention mechanisms~\cite{zou2020neural,tang2025sequential,yu2025uniform}, particularly through the transformer architecture~\cite{vaswani2017attention}.
Models like SASRec~\cite{kang2018self}, BERT4Rec~\cite{sun2019bert4rec}, and STOSA~\cite{fan2022sequential} utilize self-attention to enhance performance on user interaction data, where SASRec focuses on sequential user behavior, BERT4Rec employs bidirectional self-attention with the Cloze objective for richer feature representation, and STOSA introduces uncertainty in capturing dynamic preferences using Wasserstein Self-Attention.

These methods discussed above form the backbone of sequential recommendation and are orthogonal to our proposed approach. 
Our work also uses a transformer-based architecture as foundation. 
Ideally, our proposed method is designed to be seamlessly integrated with all of these techniques.

\subsection{Generative Recommender Systems}
Generative recommender systems have gained significant attention due to their ability to model complex user-item interactions and generate diverse, innovative recommendations. 
Early works like AutoRec~\cite{sedhain2015autorec} apply autoencoder to collaborative filtering, while models like AutoSeqRec~\cite{liu2023autoseqrec} and MAERec~\cite{ye2023graph} enhance them with incremental learning and graph representations, boosting robustness to noisy, sparse data. 
Variational Autoencoder (VAE) introduces probabilistic latent space for modeling sparse user-item interactions, with innovations like dual disentanglement modules~\cite{guo2024dualvae} and hierarchical priors~\cite{li2024hierarchical} improving interpretability and addressing sparsity.
Another popular direction is the use of Generative Adversarial Network (GAN), which enhances generative recommendation through adversarial training between generators and discriminators.
Combined with traditional collaborative filtering~\cite{dervishaj2022gan}, GAN captures complex user preferences and integrates them with techniques like Determinantal Point Processes (DPP)\cite{wu2019pd} to improve recommendation diversity. To address challenges like mode collapse, a newer GAN model~\cite{jiangzhou2024dgrm} incorporates diffusion model for stable training and reliable recommendations.

While these methods advance generative recommendation by a large margin, our research focuses on tailoring more advanced diffusion model to generate even more diverse and innovative recommendations.

\paragraph{Diffusion-Based Recommendation} 
The usage of diffusion model in sequential recommendation is still in its early stages but rapidly gaining attention due to their success in various generative tasks~\cite{ho2022video,gong2022diffuseq}.
Pioneering efforts like \cite{li2023diffurec,yang2024generate} add noise to the target item and leverage user interaction history implicitly in the reverse process. 
Based on this, Huang \textit{et al.}~\shortcite{huang2024dual} integrates both historical interactions and target items during noise addition, using both explicit sequence embeddings and implicit attention mechanisms to boost preference representation. 
Meanwhile, Wang \textit{et al.}~\shortcite{wang2024conditional} harnesses a Transformer as a conditional denoising decoder, embedding historical interactions into the model via cross-attention, thereby effectively guiding the denoising and enabling the model to focus on pertinent historical interactions.

Although these approaches are effective, they overlook the biases introduced by noise in diffusion model for recommendation task. 
In contrast, our model focuses on mitigating the distortion of user preferences caused by perturbations in both the forward and reverse diffusion processes.


\section{Preliminaries}
\label{sec:preliminaries}
This section provides a brief overview of the Flow Matching to establish the necessary background. 
Particularly, Flow Matching can be considered a simplied diffusion model, designed to construct probabilistic path between two distinct distributions, thereby enabling the transformation from simple distribution, \textit{e.g.,} a simple normal distribution \(p_n = \mathcal{N}(0, I)\), to the complex and unknown distribution, denoted as \(p_{c}\) ~\cite{esser2024scaling,nakkiran2024step}.
From the perspective of diffusion model, Flow Matching can also divide into the forward and reverse procedure. 

\paragraph{Forward Procedure} 
In the forward procedure, the model is required to map a real data point $\bm{x}_c\sim p_c$ to a noise data point $\bm{x}_n \sim p_n$. It is defined as a time-dependent flow $\phi$ as 
\begin{eqnarray}
    \phi_t(\bm{x}_c|\bm{x}_n) : \bm{x}_c \mapsto a_t \bm{x}_c + b_t \bm{x}_n 
\end{eqnarray}
where  \( t \) is a random variable uniformly sampled from the interval \([0, 1]\). 
The time-dependent hyperparameters \(a_t\) and \(b_t\) follow two constraints: if \(a_0 = 1\), then \(b_0 = 0\), ensuring \(\phi_0(\bm{x}_c|\bm{x}_n) = \bm{x}_c\) at \(t = 0\), and if \(a_1 = 0\), then \(b_1 = 1\), ensuring \(\phi_1(\bm{x}_c|\bm{x}_n) = \bm{x}_n\) at \(t = 1\). 
These ensure a transition from the target distribution \(\bm{x}_c\) to the normal distribution \(\bm{x}_n\) over time. 
$\phi_t(\bm{x}_c|\bm{x}_n)$ provides a concise representation of the Flow Trajectory, illustrating the manner in which states change during the Flow process.
To simplify the notation, we denote the intermediate variable perturbed by noise as 
\begin{equation}
\label{euq:vector}
    \bm{z}_t = a_t \bm{x}_c + b_t \bm{x}_n.
\end{equation}

To characterize the flow \(\phi_t(\cdot | \bm{x}_n)\), a vector field \(u_t\) is employed to construct the time-dependent diffeomorphic map \(\phi_t\) as follows:
\begin{equation}
\label{euq:u_t}
\bm{u}_t(\bm{z}_t | \bm{x}_n) := \phi_t'(\phi_t^{-1}(\bm{z}_t | \bm{x}_n) | \bm{x}_n),
\end{equation}
where \(\phi_t^{-1}(\bm{z}_t | \bm{x}_n)\) represents the inverse function, which computes \(\bm{x}_c\) based on the perturbed noise. The notation \(\phi_t'\) denotes the differential of \(\phi_t\).

The objective of the training process is to learn and predict this vector field with a $\Theta$ \textbf{parameterized vector field} $\bm{v}_\Theta(\bm{z}_t, t)$ as 
\begin{equation}
\label{equ:Loss}
    \mathcal{L}_{FM} = \mathbb{E}_{t, p_t(\bm{z}|\bm{x}_c), p(\bm{x}_c)} ||\bm{v}_\Theta(\bm{z}_t, t) - u_t(\bm{z}_t|\bm{x}_n)||_2^2, 
\end{equation}
where $||\cdot||_2^2$ is the L2-norm.

\paragraph{Reverse Procedure} 
In the reverse process, \textit{i.e.}, the inference procedure, the model reconstructs \(\bm{x}_c\) by solving the following ordinary differential equation~(ODE):
\[
   d\bm{z}_t =  -\bm{v}_\Theta(\bm{z}_t, t) \, dt,
\]
where $\bm{z}_t$ is the linear combination of $\bm{x}_c$ and $\bm{x}_n$. 
In this work, we employ a \textit{deterministic reverse sampler}, \textit{i.e.}, the Euler method, to solve this ODE.

Remarkably, the diffusion-based recommendation models ~\cite{li2023diffurec} and~\cite{wang2023diffusion}, typically employ an SDE-based \textit{stochastic reverse sampler}, which introduces the stochastic disturbances, \textit{i.e.}, the variance, to the reverse diffusion process. 
However, this stochasticity deviates from the objective of sequential recommendation, potentially resulting in irrelevant predictions when attempting to accurately identify a user's next interaction item, which ultimately undermines the user experience.
Contrarily, the ODE-based \textit{deterministic reverse sampler} does not introduce any random perturbations during the generation process, ensuring that the generation results meet the user's personalized preferences.

\section{Methodology}
\label{sec:methodology}

\subsection{Problem Statement}

For the sequential recommendation task, we define a set of users $\mathcal{U} = \{u_1, u_2, \dots, u_{|\mathcal{U}|}\}$ and a set of items $\mathcal{I} = \{i_1, i_2, \dots, i_{|\mathcal{I}|}\}$, where $|\mathcal{U}|$ and $|\mathcal{I}|$ represent the total number of users and items, respectively. 
Each user $u \in \mathcal{U}$ has an interaction history represented as a chronological sequence of items $\mathcal{S} = (i_1, i_2, \dots, i_m)$, where $i_m$ corresponds to the $m$-th item interacted with by the user. Here, $m$ is the length of the interaction sequence. Formally, we aim at generating the next recommendation $i_{m+1}$, maximizing a specific metric $\Theta$ as
\begin{equation}
i_{m+1} = \mathop{\arg\max}_{i_{m+1}} \\ \Theta(i_{m+1}|\mathcal{S}).
\end{equation}




\begin{figure*}[htbp]
    \centering
    \includegraphics[width=0.94\textwidth]{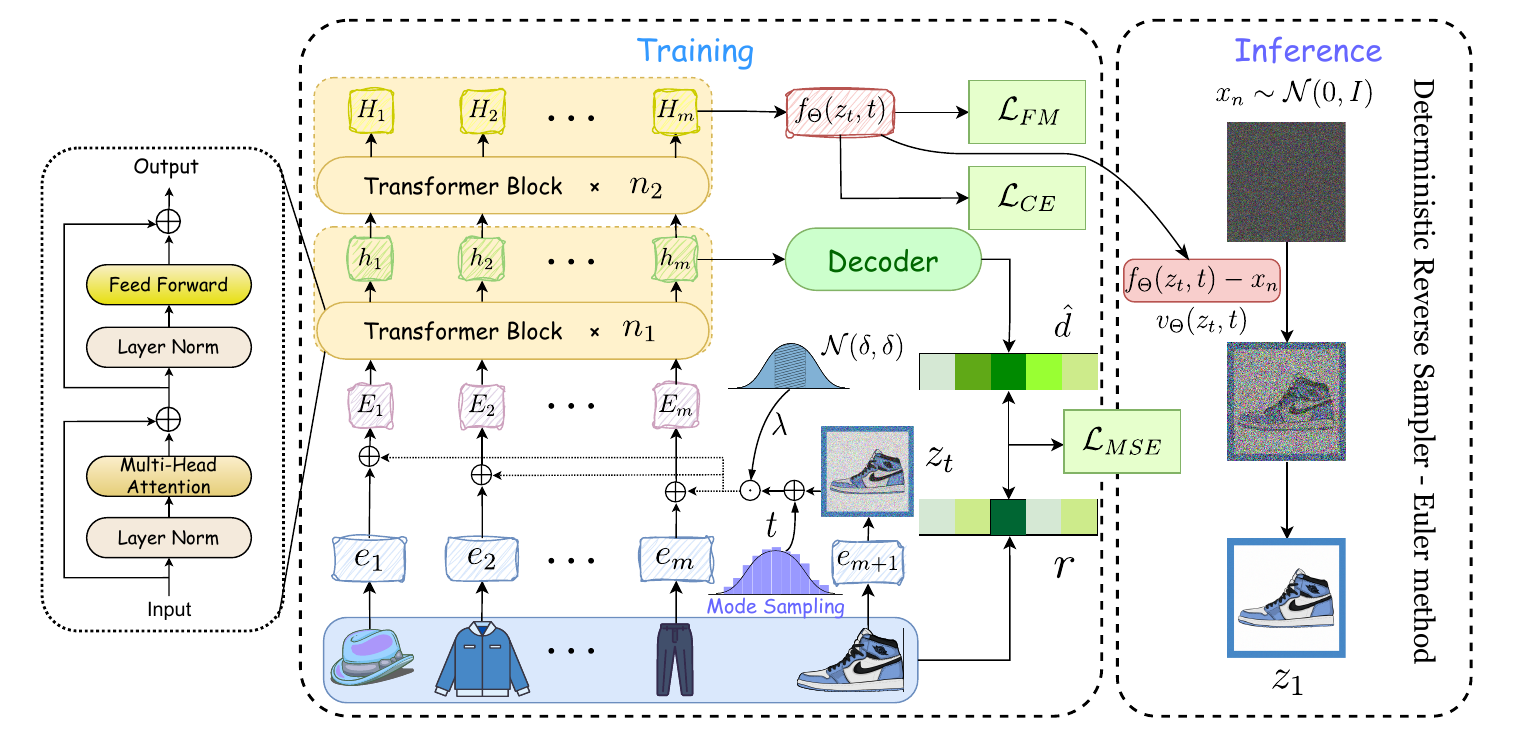}
    \vspace{-0.2cm}
    \caption{The framework of the \name.
    In the training process, our design incorporates the development of straight flow trajectories, modifications to the learning target \(\mathcal{L}_{FM}\), the design of a decoder-based model, and the implementation of regularized loss functions \(\mathcal{L}_{CE}\) and \(\mathcal{L}_{MSE}\). In the inference process, we present a deterministic reverse sampler that generates recommendations.
    }
    \label{fig:flowmatching}
    \vspace{-0.4cm}
\end{figure*}

\subsection{Flow Matching based Sequential Recommender Model}
\label{subsec:FM method}
This section offers a detailed explanation of both the forward and reverse processes in \name. 
In the forward process, our designs include the development of flow trajectories, modification of the learning target, design of the parameterized vector field, and the corresponding learning loss.
In the reverse process, we discuss the specific selection and implementation of the reverse sampler.

\subsubsection{Forward Process}



In the context of sequence recommendation task, we adopt the target distribution \( p_c \) in Flow Matching as follows:
\begin{equation}
    p(\bm{x}_c | \bm{e}_{m+1}) = \mathcal{N}(\bm{x}_c; \bm{e}_{m+1}, 0),
\end{equation}
where \( \bm{e}_{m+1} \) represents the embedding of the target item, generated using the equation:
\begin{equation}
    \bm{e}_{m+1}= \texttt{Embedding}(i_{m+1}).
\end{equation}
Here, \texttt{Embedding} denotes an embedding module that transforms discrete next item ID $i_{m+1}$ into dense vector representation. 
Notably, the parameter \(\bm{e}_{m+1}\) is trainable, which is different from the traditional Flow Matching settings.

\paragraph{Straight Trajectory Flow and Modified Loss}
The time-dependent hyper-parameters \( a_t \) and \( b_t \) in the forward process define the trajectory of the generated flow, enabling the flexible selection of flow paths to control the flow process. The Straight Trajectory, characterized by \( a_t = (1-t) \) and \( b_t = t \), is known for its simplicity and computational efficiency, making it widely employed in various studies \cite{lipman2022flow,liu2022flow}. Following this setting, we define the time-dependent flow of the next item recommendation as follows:
\begin{equation}
   \bm{z}_t = (1-t) \bm{x}_c + t \bm{x}_n, \quad \text{where } \bm{x}_n \sim \mathcal{N}(0, I), 
\end{equation}
where \( I \) denotes the identity matrix.
Further discussions regarding its effectiveness are provided in Section~\ref{subsec: analysis experiment}. 
The variable $t$ is sampled using the Mode Sampling with Heavy Tails~\cite{esser2024scaling} method as 
\begin{equation}
t = g(k; s) = 1 - k - s \times \left( \cos^2 \left( \frac{\pi}{2} k \right) - 1 + k \right),
\end{equation}
where the parameter $s$ represents a scaling factor that governs the density distribution of the time step sampling. $k \in \mathcal{U}(0, 1)$ is a random variable. 
Further discussions on the sampling methods and the impact of the parameter $s$ on model performance are provided in Appendix~\ref{sec:experienment_appendix}.


To characterize the flow \(\phi_t(\cdot | \bm{x}_n)\), a vector field can be defined in the following form:
\begin{equation}
\label{eq:u_t_rf}
    \bm{u}_t(\bm{z}_t | \bm{x}_n) = -\phi_t^{-1}(\bm{z}_t | \bm{x}_n) + \bm{x}_n,
\end{equation}
where \( \phi_t^{-1}(\bm{z}_t | \bm{x}_n) \) represents the inverse of the flow function at time \( t \), conditioned on \( \bm{x}_n \). By substituting into the Flow Matching loss defined in Equation~(\ref{equ:Loss}), we can reformulate the training loss as follows: 
\begin{small}
\begin{flalign}
\label{equ:navie_fm_loss}
\mathcal{L}_{FM} &= \mathbb{E}_{t, p_t(\bm{z}|\bm{x}_c), p(\bm{x}_c)}||\bm{v}_\Theta(\bm{z}_t, t)  - (-~\phi_t^{-1}(\bm{z}_t | \bm{x}_n)+\bm{x}_n)||_2^2  \\ 
&\label{eqn_loss:2} = \mathbb{E}_{t, p_t(\bm{z}|\bm{x}_c), p(\bm{x}_c)}|| (-f_\Theta(\bm{z}_t, t) + \bm{x}_n)  - (-\bm{x}_c + \bm{x}_n)||_2^2 \\
&= \mathbb{E}_{t, p_t(\bm{z}|\bm{x}_c), p(\bm{x}_c)}||f_\Theta(\bm{z}_t, t) -~\bm{x}_c||_2^2 \label{equ:loss_x1}.
\end{flalign}
\end{small}
\noindent The first derivation relies on the fact $\bm{x}_n = \phi_t^{-1}(\bm{z}_t | \bm{x}_n)$. 
The second equation replaces the learning target $\bm{v}_\Theta(\bm{z}_t, t)$ with $(-f_\Theta(\bm{z}_t, t) + \bm{x}_n)$, leading to the modified target in Equation~(\ref{equ:loss_x1}). 
The reason for replacing the learning target is that predicting the intricate combination of real data and noise can be difficult and detrimental to recommendation systems, leading to the recommendation of diverse but irrelevant items.

\paragraph{Parameterized Vector Field}


To parameterize \( f_\Theta(\bm{z}_t, t) \) for predicting real data, we first integrate the historical interaction sequence with the noised data and then model the noised historical interactions using a robust transformer decoder. Specifically, we combine the historical interaction sequence with the noised data as follows~\cite{li2023diffurec}:
\begin{eqnarray}
    \bm{E}_i(\bm{z}_t, t) = \bm{e}_i + \lambda_{i} \odot (\bm{z}_t + t), \quad \lambda_i \sim \mathcal{N}(\delta, \delta)
\end{eqnarray}
where \( \odot \) denotes element-wise multiplication, \( \lambda_{i} \) is sampled from a Gaussian distribution, and \( \delta \) is a hyperparameter representing both the mean and variance of the distribution. The term \( \lambda_{i} \) is instrumental in balancing the fusion ratio between the historical interaction sequence and the noised data.

Afterwards, we employ a decoder, denoted as $\texttt{Decoder}_1$, which consists of \( n_1 \) layers of unidirectional self-attention based transformer, to obtain the hidden states as follows:
\begin{eqnarray}
\label{equ:h_n}
    [\bm{h}_1, \ldots, \bm{h}_m] = \texttt{Decoder}_1([\bm{E}_1(\bm{z}_t, t), \ldots, \bm{E}_m(\bm{z}_t, t)]),
\end{eqnarray}
where \( \bm{h}_1 \) to \( \bm{h}_m \) are the intermediate outputs that will be utilized for reconstructing the interaction matrix and calculating the reconstruction loss. 
Next, we model \( f_\Theta(\bm{z}_t, t) \) by taking the \texttt{LAST} token output from an additional \( n_2 \)-layer decoder, denoted as $\texttt{Decoder}_2$, as follows:
\begin{eqnarray}
\label{equ:f_theta}
    f_\Theta(\bm{z}_t, t) = \texttt{LAST}(\texttt{Decoder}_2([\bm{h}_1, \ldots, \bm{h}_m])).
\end{eqnarray}

\paragraph{Regularized Losses}

The modified Flow Matching loss presented in Equation~(\ref{equ:loss_x1}) is not entirely suitable for recommendation task because the vector \(\bm{x}_c\) is trainable rather than fixed. This might lead to the problem that different \(\bm{x}_c\) converge to a single embedding, which is obviously meaningless. 
To mitigate this issue, we introduce a cross-entropy loss $\mathcal{L}_{CE}$ as a regularizer that differentiates between various item embeddings, thereby preventing the aforementioned problem as:
\begin{eqnarray}
\label{equ:loss_ce}
\mathcal{L}_{CE} &=&  -\log \hat{y}_{m+1} \\
\hat{y}_{m+1} &=& \frac{\exp(f_\Theta(\bm{z}_t, t) \cdot \bm{e}_{m+1})}{\sum_{i \in \mathcal{I}} \exp(f_\Theta(\bm{z}_t, t) \cdot \bm{e}_i)}.
\end{eqnarray}
where $\hat{y}_{m+1}$ is the normalized score of recommending $i_{m+1}$.


Furthermore, to mitigate the detrimental effects of noise perturbation on model performance, we incorporate an additional reconstruction loss that interprets the user's history from the final token of the hidden representation \( \bm{h}_m \) as follows:
\begin{equation}
\bm{\hat{d}} = \texttt{Decoder}_{\omega}(\bm{h}_{m}),
\end{equation}
where $\bm{\hat{d}}\in \mathbb{R}^{|\mathcal{I}|}$ is the predicted  user-item interaction information. \(\omega \) represents the parameters of the MLP-based $\texttt{Decoder}_{\omega}$. 
We optimize these parameters using a Mean Squared Error (MSE) loss, enabling the generation of more accurate and personalized recommendations as follows:
\begin{equation}
\mathcal{L}_{MSE} = \| \bm{\hat{d}} - \bm{r} \|^2,
\end{equation}
where \( \bm{r} \in \mathbb{R}^{|\mathcal{I}|} \) is a binary vector, with 1 indicating an interaction between the user and the item, and 0 indicating no interaction.

Finally, the loss function utilized during training is formulated by integrating three distinct components, represented as follows:
\begin{eqnarray}
    \mathcal{L} = \mathcal{L}_{FM} +  \alpha  \mathcal{L}_{CE} + \beta \mathcal{L}_{MSE},
\end{eqnarray}
where \(\alpha\) and \(\beta\) are hyperparameters that govern their relative importance. 
A more detailed training procedure of \name is presented in Appendix~\ref{sec:Algorithm}.





\subsubsection{Reverse Process}
In the reverse process, we employ a \textit{deterministic reverse sampler} to model the generative process, thereby mitigating the errors introduced by the \textit{stochastic reverse sampler} in the diffusion-based recommendation system. Specifically, the \textit{deterministic reverse sampler} is defined as the updating function for the following ordinary differential equation: 
\begin{equation}
\label{equ:ODE}
    d\bm{z}_t = - u_t(\bm{z}_t|\bm{x}_n) dt = (f_\Theta(\bm{z}_t, t) - \bm{x}_n) dt.
\end{equation}
To solve this equation, we iteratively apply the Euler method to compute the point transformations guided by the vector field, as given by 
\begin{equation}
    \bm{z}_{t+\Delta t} = \bm{z}_t + (f_\Theta(\bm{z}_t, t) - \bm{x}_n) \Delta t,
\end{equation}
where $\Delta t$ is determined by a custom number of Euler method steps, denoted as \(q\), with \(\Delta t = \frac{1}{q}\). This process is iteratively calculated from \(t = 0\) to \(t = 1\), resulting in \(\bm{z}_1\) as the fully denoised generation \(\bm{\hat{x}_c}\), which represents the corresponding item embedding. The inference procedure of \name is presented in Appendix~\ref{sec:Algorithm}.

\section{Experiment}
\label{sec:experiment}
This section presents comprehensive experiments to demonstrate the effectiveness of \name.

\subsection{Experimental Settings}
\paragraph{Dataset}

\begin{table}
    \centering
    \resizebox{0.43\textwidth}{!}{%
        \begin{tabular}{lrrrrr}
            \toprule
            \textbf{Dataset} & \textbf{\# Users} & \textbf{\# Items} & \textbf{\# Actions} & \textbf{Sparsity} \\
            \midrule
            Beauty & 22,361 & 12,101 & 198,502 & 99.93\% \\
            Steam & 281,428 & 13,044 & 3,485,022 & 99.90\% \\
            Movielens-100k & 943 & 1,682 & 100,000 & 93.70\% \\
            Yelp & 28,298 & 59,951 & 1,764,589 & 99.90\% \\
            \bottomrule
        \end{tabular}
    }
    \vspace{-0.2cm}
    \caption{The statistics of the datasets.}
    \vspace{-0.4cm}
    \label{tab:dataset}
\end{table}
\begin{table*}[htbp]
    \centering
    \resizebox{0.95\textwidth}{!}{%
    \begin{tabular}{llrrrrrrrrrrrr}
        \toprule
        \textbf{Dataset} & \textbf{Metric} & \textbf{GRU4Rec}& \textbf{Caser}& \textbf{SASRec}& \textbf{BERT4Rec}& \textbf{STOSA}& \textbf{AutoSeqRec}& \textbf{DreamRec}& \textbf{DiffuRec}& \textbf{FM4Rec} & \textbf{$\Delta \%$} \\
        \midrule
        \multirow{6}{*}{%
            \makebox[1.5cm]{%
                \rotatebox[origin=c]{90}{\textbf{Beauty}}%
            }%
        }
        & HR@5    & 1.0112 & 1.6188 & 3.2688 & 2.1326 & \text{3.8814} & \text{4.9628} & \text{4.9833} & \underline{\text{5.3943}}& \textbf{5.8373} & 8.21\% \\
        & HR@10  & \text{1.9370} & \text{2.8166} & \text{6.2648} & \text{3.7160} & \text{6.1262} & \text{7.1016} & \text{6.9821} & \underline{\text{7.8374}} & \textbf{8.2693} & 5.51\% \\
        & HR@20 & \text{3.8531} & \text{4.4048} & \text{8.9791} & \text{5.7922} & \text{9.0954} & \text{9.3342} & \text{9.4531} & \underline{\text{10.9358}} & \textbf{11.626} & 6.31\%  \\
        & NDCG@5 & \text{0.6084} & \text{0.9758} & \text{2.3989} & \text{1.3207} & \text{2.4859} & \text{3.3186} & \text{3.2507} & \underline{\text{3.9153}} & \textbf{4.1631} & 6.33\% \\
        & NDCG@10 & \text{0.9029} & \text{1.3602} & \text{3.2305} & \text{1.8291} & \text{3.2053} & \text{4.0157} & \text{3.9769} & \underline{\text{4.6971}} & \textbf{4.9461} & 5.30\% \\
        & NDCG@20  & \text{1.3804} & \text{1.7595} & \text{3.6563} & \text{2.3541} & \text{3.9491} & \text{5.0133} & \text{4.9860} & \underline{\text{5.4784}} & \textbf{5.7876} & 5.64\%  \\
        \midrule
        \multirow{6}{*}{%
            \makebox[1.5cm]{%
                \rotatebox[origin=c]{90}{\textbf{Steam}}%
            }%
        }
        & HR@5    & \text{3.0124} & \text{3.6053} & \text{4.7428} & \text{4.7391} &\text{4.8546} & \text{5.0021} & \text{5.1267} & \underline{\text{6.0073}} & \textbf{6.5254} & 8.62\% \\
        & HR@10  & \text{5.4257} & \text{6.4940} & \text{8.3763} & \text{7.9448} & \text{8.5870} & \text{8.7741} & \text{8.9875} & \underline{\text{9.8437}} & \textbf{10.5908} & 7.59\% \\
        & HR@20 & \text{9.2319} & \text{10.9633} & \text{13.6060} & \text{12.7322} & \text{14.1107} & \text{14.6752} & \text{15.0871} & \underline{\text{15.3817}} & \textbf{16.4669} & 7.06\%  \\
        & NDCG@5 & \text{1.8293} & \text{2.1586} & \text{2.8842} & \text{2.9708} & \text{2.9220} & \text{3.0912} & \text{3.1507} & \underline{\text{3.8109}} & \textbf{4.1878} & 9.89\% \\
        & NDCG@10 & \text{2.6033} & \text{3.0846} & \text{4.0489} & \text{4.0002} & \text{4.1191} & \text{4.4729} & \text{4.6416} & \underline{\text{5.0429}} & \textbf{5.4925} & 8.91\% \\
        & NDCG@20 & \text{3.5572} & \text{4.2043} & \text{5.3630} & \text{5.2027} & \text{5.5072} & \text{5.9823} & \text{5.9701} & \underline{\text{6.4340}} & \textbf{6.9689} & 8.31\%  \\
        \midrule
        \multirow{6}{*}[1pt]{%
            \makebox[1.5cm]{%
                \rotatebox[origin=c]{90}{\textbf{Movielens-100k}}%
            }%
        }
        & HR@5    & \text{7.7412} & \text{6.0438} & \text{6.5748} & \text{5.0901} &\text{8.0148} & \underline{\text{8.7320}} & \text{7.3692} & \text{7.5209} & \textbf{9.3338} & 6.89\% \\
        & HR@10 & \text{12.1951} & \text{11.2426} & \text{13.5737} & \text{9.3319} & \text{13.6542} & \underline{\text{14.6641}} & \text{12.4377} & \text{12.8501} & \textbf{15.4934} & 5.66\% \\
        & HR@20 & \text{21.8451} & \text{19.5189} & \text{22.6935} & \text{16.8611} & \text{21.7761} & \underline{\text{22.8724}} & \text{20.8357} & \text{19.4127} & \textbf{24.3146} & 6.31\%  \\
        & NDCG@5 & \text{4.5982} & \text{3.3721} & \text{4.1333} & \text{3.0850} & \text{4.9721} & \underline{\text{5.5010}} & \text{4.2503} & \text{4.6969} & \textbf{5.6848} & 3.34\% \\
        & NDCG@10& \text{6.0326} & \text{5.0683} & \text{6.3427} & \text{4.4568} & \text{5.2159} & \underline{\text{7.3955}} & \text{5.9837} & \text{6.4136} & \textbf{7.6571} & 3.53\% \\
        & NDCG@20& \text{8.4727} & \text{7.1439} & \text{8.6340} & \text{6.3442} & \text{8.3302} & \underline{\text{9.4584}} & \text{7.8234} & \text{8.0459} & \textbf{9.8158} & 3.78\%  \\
        \midrule
        \multirow{6}{*}[1pt]{%
            \makebox[1.5cm]{%
                \rotatebox[origin=c]{90}{\textbf{Yelp}}%
            }%
        }
        & HR@5    & \text{2.4560} & \text{2.0956} & \text{2.8389} & \text{2.2465} &\text{1.9360} & \text{OOM} & \text{1.7486} & \underline{\text{3.0390}} & \textbf{3.3084} & 8.86\% \\
        & HR@10  & \text{4.2335} & \text{3.7140} & \text{4.8569} & \text{4.0581} & \text{3.3858} & \text{OOM} & \text{1.9362} & \underline{\text{5.075}} & \textbf{5.4421} & 7.23\% \\
        & HR@20 & \text{7.4952} & \text{6.6189} & \text{8.2656} & \text{7.0433} & \text{5.7285} & \text{OOM} & \text{3.5873} & \underline{\text{8.5447}} & \textbf{9.0631} & 6.07\%  \\
        & NDCG@5 & \text{1.5588} & \text{1.3108} & \text{1.8301} & \text{1.4027} & \text{1.2100} & \text{OOM} & \text{1.1740} & \underline{\text{1.9868}} & \textbf{2.1174} & 6.57\% \\
        & NDCG@10 & \text{2.1269} & \text{1.8311} & \text{2.5466} & \text{1.9732} & \text{1.6728} & \text{OOM} & \text{1.5268} & \underline{\text{2.6352}} & \textbf{2.7855} & 5.71\% \\
        & NDCG@20 & \text{2.9431} & \text{2.5607} & \text{3.3144} & \text{2.7233} & \text{2.2584} & \text{OOM} & \text{2.3641} & \underline{\text{3.4395}} & \textbf{3.618} & 5.19\%  \\
        \bottomrule
    \end{tabular}%
    }
    \vspace{-0.2cm}
     \caption{Overall performance comparison across four benchmark datasets, presented as percentages (\%). We highlight the highest-performing metric values in \textbf{bold} and the second-best values in \underline{underlined}. The symbol $\Delta$ indicates the relative performance improvement of \name compared to the best baseline model. OOM refers to the out-of-memory problem.}
     \vspace{-0.2cm}
    \label{tab:result}
\end{table*}
We evaluate \name's effectiveness using four widely recognized publicly available datasets: \textbf{(1)} \textit{Amazon Beauty}~\cite{ni2019justifying} contains global purchasing interactions and user reviews for beauty products on the Amazon platform, documenting the purchasing history of 22,361 users across 12,101 products.
\textbf{(2)} \textit{Steam} is a leading PC game distribution platform with 3,480,000 interaction records from 280,000 gamers; 
\textbf{(3)} \textit{Movielens-100k}~\cite{harper2015movielens} is a widely used benchmark dataset in sequential recommendation research, providing ratings from 943 users on 1,682 movies from the Movielens platform; and 
\textbf{(4)} \textit{Yelp} is a popular review site featuring user reviews and 1,764,589 ratings for various businesses, including restaurants, entertainment venues, and hotels. The statistics of the dataset are provided in Table~\ref{tab:dataset}.

\paragraph{Baselines}
We compare \name against both widely adopted and recently introduced baseline models: \textbf{(1) GRU4Rec} utilizes GRU to capture in-session behavioral patterns and predict subsequent user item preferences~\cite{hidasi2015session}; \textbf{(2) Caser} leverages convolutional neural networks to map user action sequences into both temporal and latent spaces~\cite{tang2018personalized}; \textbf{(3) SASRec} introduces a decoder architecture for sequential recommender model that effectively captures long-term dependencies in user behavior~\cite{kang2018self}; \textbf{(4) BERT4Rec} employs a bidirectional Transformer architecture, coupled with a Cloze task for training, to effectively learn users' dynamic preferences~\cite{sun2019bert4rec}; \textbf{(5) STOSA} utilizes Wasserstein attention to introduce a degree of uncertainty within its model, allowing for the accurate representation of evolving user preferences~\cite{fan2022sequential}; \textbf{(6) AutoSeqRec} leverages a multi-autoencoder framework to fuse long-term user preferences and short-term interests for sequential recommendation~\cite{liu2023autoseqrec}; \textbf{(7)~DreamRec} achieves direct generation of personalized oracle item embeddings through guided diffusion model~\cite{yang2024generate}; \textbf{(8)~DiffuRec} models items as distributions using diffusion model, capturing multi-faceted content and user preferences~\cite{li2023diffurec}.

\paragraph{Evaluation Protocol} 
Following the procedures in \cite{sun2019bert4rec,li2023diffurec}, we split user interaction sequences into three parts: the first \( m-2 \) sequences formed the training set, while \( i_{m-1} \) and \( i_m \) served as targets for the validation and test sets, respectively. We evaluated performance using the metrics HR@$K$ (Hit Rate@$K$) and NDCG@$K$ (Normalized Discounted Cumulative Gain@$K$). Each baseline model generates a ranked list of items predicted for the next interaction based on user history, considering all dataset items as candidates, with \( K \) values set at \( \{5, 10, 20\} \).


\paragraph{Implementation Details}
The implementation details are as follows: both $\texttt{Decoder}_1$ and $\texttt{Decoder}_2$ consist of 2 layers, with a hidden dimension of 128 and 4 attention heads. The item embedding dimension is also set to 128. The $\texttt{Decoder}_{w}$ is a three-layer MLP with tanh activations, featuring layer sizes of $\{128, 512, 2048, |\mathcal{I}|\}$, mapping 128-dimensional decoder outputs to the number of items. Hyperparameters include a batch size of 512, a learning rate of 0.001, and a maximum user interaction sequence length of 50. The loss weighting parameters $\alpha$ and $\beta$ are set to 0.2 and 0.4, respectively. The scaling parameter $s$ in the timestep schedule is set to 1.0. Besides, we use 30 Euler integration steps for generation. All experiments are conducted on a server with two Intel XEON 6271C processors, 256 GB of memory, and four NVIDIA RTX 3090 Ti GPUs.

\subsection{Overall Performance Comparison}
The experimental results and performance comparisons with baseline models are presented in detail in Table~\ref{tab:result}. 
From the table, we have the following observations: \textbf{(1) The \name model shows a notable advantage over existing SOTA methods across four benchmark datasets.} Significant improvements are consistently observed in the Beauty, Yelp, Movielens-100k, and Steam datasets with HR increasing by $5.5-8.9\%$ and NDCG rising by $3.3-9.9\%$; 
\textbf{(2)~\name can effectively eliminate the negative influence of noise perturbations associated with diffusion-based recommendation methods.} Specifically, we observe that \name outperforms both DreamRec and DiffuRec, achieving superior results with an average improvement of 12.44\% on HR@$5$, which confirms that the \textit{deterministic reverse sampler} and regularized loss are beneficial for generating more accurate recommendation.
\textbf{(3)~Generative models demonstrate greater effectiveness in capturing user preferences and providing enhanced recommendations.} Particularly, DreamRec, DiffuRec, and \name show marked improvements in performance over SOTA traditional sequential recommender models by large margins of 13.7\%, 27.93\%, and 34.98\% on HR@$5$.

\subsection{Analytical Experiment}
\label{subsec: analysis experiment}
This section evaluates the effectiveness of each design option in \name and analyzes the impact of hyperparameter configurations on model performance. We present results of \textit{Beauty} and \textit{Movielens-100k} as examples. Additional results are available in Appendix~\ref{sec:experienment_appendix}.

\begin{table}
    \centering
    
    \resizebox{0.48\textwidth}{!}{%
        \begin{tabular}{lrrrrrr}
            \toprule
            \multicolumn{7}{c}{\textit{Beauty}}  \\ \hline\hline 
             & \textbf{HR@5} & \textbf{HR@10} & \textbf{HR@20} & \textbf{NDCG@5} & \textbf{NDCG@10} & \textbf{NDCG@20} \\
            \hline 
            $v$-prediction & 0.5753 & 1.1962 & 2.0276 & 0.3138 & 0.5104 & 0.7948 \\
            \name  & \textbf{5.8373} & \textbf{8.2693} & \textbf{11.626} & \textbf{4.1631} & \textbf{4.9461} & \textbf{5.7876} \\
            \toprule 
            \multicolumn{7}{c}{\textit{Movielens-100k}}  \\ \hline\hline
             & \textbf{HR@5} & \textbf{HR@10} & \textbf{HR@20} & \textbf{NDCG@5} & \textbf{NDCG@10} & \textbf{NDCG@20} \\
            \midrule
            $v$-prediction & 0.8251 & 1.5239 & 2.2085 & 0.4637 & 0.7524 & 0.9896 \\
            \name & \textbf{9.3338} & \textbf{15.4934} & \textbf{24.3146} & \textbf{5.6848} & \textbf{7.6571} & \textbf{9.8158} \\
            \bottomrule
        \end{tabular}
    }
    \vspace{-0.2cm}
    \caption{Performance comparison using different Flow Matching losses, presented as percentages (\%). The $v$-prediction approach employs the naive training loss. \name utilizes the modified Flow Matching loss.}
    \vspace{-0.2cm}
    \label{tab:prediction}
\end{table}

\paragraph{Influence of Flow Matching Loss}
\label{par:loss}
In this work, we have modified the Flow Matching loss from directing predicting the overall vector field, \textit{i.e.}, Equation~(\ref{equ:navie_fm_loss}), to the modified vector field, \textit{i.e.}, Equation~(\ref{equ:loss_x1}). 
To demonstrate its effectiveness, Table~\ref{tab:prediction} presents the comparison between the model using the modified loss and the naive loss~($v$-prediction in Table~\ref{tab:prediction}) from the Flow Matching loss. 
From the table, we observe a notable performance drop when using the naive Flow Matching loss. This suggests that \textbf{directly predicting the vector field results in inaccurate modeling of user preferences}, which significantly undermines the model's performance.


\begin{figure}[ht]
    \centering
    \includegraphics[width=0.45 \textwidth]{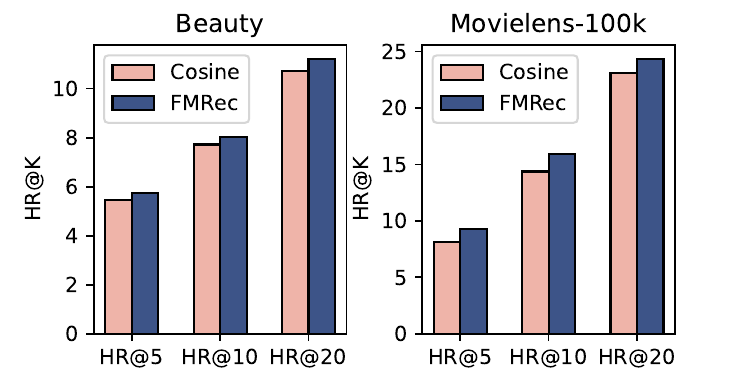}
    \vspace{-0.2cm}
    \caption{Performance comparison based on different flow trajectories, measured as percentages(\%): ``Cosine'' represents the results obtained using the Cosine trajectory, while ``\name'' denotes the use of the straight trajectory.}
    \vspace{-0.2cm}
    \label{fig:flow_trajectory}
\end{figure}

\paragraph{Effectiveness of Straight Trajectories}
To demonstrate the effectiveness of straight trajectories, we use the Cosine trajectory as a baseline for comparison. This trajectory has been employed in IDDPM~\cite{nichol2021improved} to achieve superior performance compared to DDPM~\cite{ho2020denoising}. 
In Figure~\ref{fig:flow_trajectory}, we present the results for both the Cosine trajectory (denoted as ``Cosine'') and straight trajectories (\name) on the \textit{Beauty} and \textit{Movielens-100k} datasets. 
The figure reveals a noticeable performance drop when using the Cosine trajectory, highlighting that straight trajectories improve robustness against error propagation. 
This robustness facilitates faster convergence to optimal results with fewer iterations.



\begin{figure}[htbp]
    \centering
    \includegraphics[width=0.45 \textwidth]{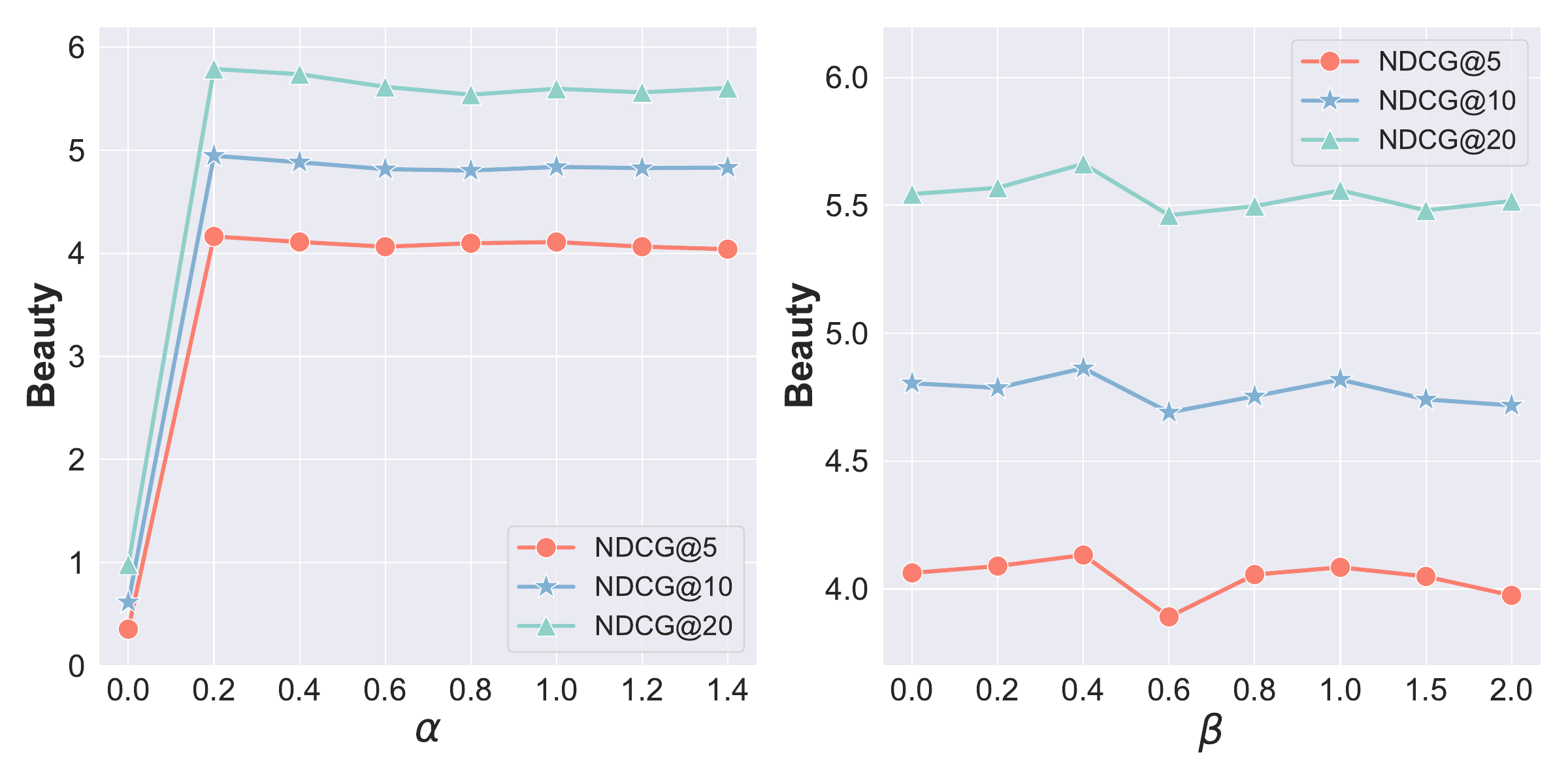}
    \vspace{-0.3cm}
    \caption{Comparison of model performance across various $\alpha$ and $\beta$ on the \textit{Beauty} dataset, measured as percentages(\%). These parameters affect the importance of the cross-entropy loss and the reconstruction loss.}
    \vspace{-0.2cm}
    \label{fig:loss_weight}
\end{figure}

\paragraph{Effectiveness of Regularized Lossesd}
\name incorporates two regularized loss functions to enhance the model's ability for recommendation. 
This section analyzes its effectiveness by examining the influence of the parameters $\alpha$ and $\beta$, as illustrated in Figure~\ref{fig:loss_weight}. 
Specifically, the figure displays the model's performance when trained with various combinations of $\alpha$ and $\beta$ on the \textit{Beauty} dataset. 
From Figure~\ref{fig:loss_weight}, we observe that an optimal setting of $\alpha = 0.2$ and $\beta = 0.4$ yields the best performance. 
Low values of $\alpha$ can significantly impact model performance, while excessively high values of $\alpha$ can also adversely affect it to some extent. In contrast, choosing an appropriate $\beta$ can enhance the overall performance of the model.
These results indicate that \textbf{both regularized loss functions are crucial for maintaining the effectiveness of the proposed \name model.}


\begin{figure}[htbp]
    \centering
    \includegraphics[width=0.48 \textwidth]{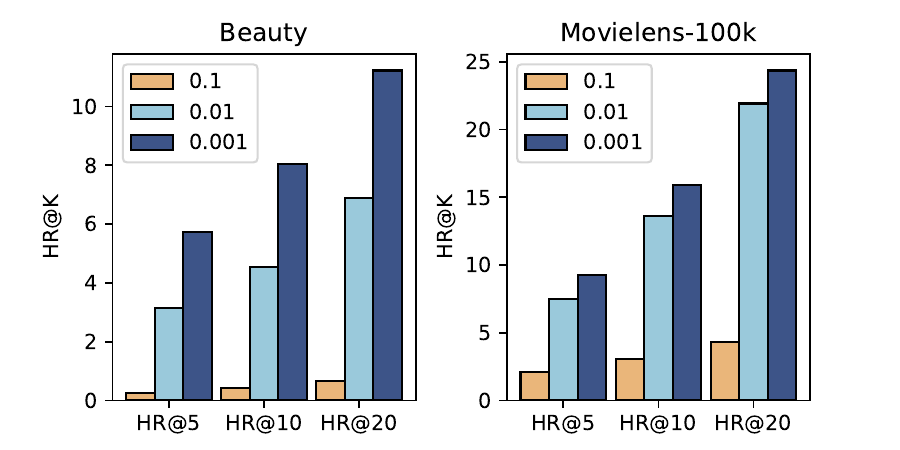}
    \vspace{-0.5cm}
    \caption{Performance comparison across different values of $\delta$, measured as percentages(\%), where $\delta$ controls the weight of noise perturbation in the user's sequential interactions fed into the model.}
    \vspace{-0.2cm}
    \label{fig:Hyperparameter delta}
\end{figure}

\paragraph{Influence of Noise Perturbation}
Before feeding the user's sequential interactions into the model, \name incorporates noise to meet the requirements of Flow Matching, where $\delta$ controls the fusion ratio's influence. 
To examine the impact of noise perturbation, we conduct experiments with $\delta \in \{0.1, 0.01, 0.001\}$. 
Figure~\ref{fig:Hyperparameter delta} presents the performance comparison across different $\delta$ values.
From the figure, we observe that the best performance is achieved when $\delta$ was set to $0.001$, demonstrating that an appropriate level of noise perturbation enhances performance.
Conversely, too large noise perturbation might lead to deviations in the calculation of the vector field during the reverse process and negatively impact the generation performance.

\section{Conclusion}

In conclusion, this work introduces a sequential recommendation model using Flow Matching, a simplified diffusion model that effectively mitigates noise perturbations of the diffusion-based model. By employing a straight flow trajectory and deriving a noise-free training target, our method reduces error accumulation in modeling user preferences. Additionally, the integration of a decoder architecture with an interaction reconstruction loss increases robustness against noise, ensuring precise user preference modeling. Furthermore, our deterministic reverse sampler, utilizing the Euler method, removes random perturbations during recommendation generation. Extensive experiments on four benchmark datasets demonstrate that our method, \name, achieves an average improvement of \textbf{6.53\%} over SOTA techniques.

\section*{Acknowledgment}
We express our sincere gratitude for the financial support provided by the National Natural Science Foundation of China (No. U23A20305, NO. {62302345}), the {CCF-ALIMAMA TECH Kangaroo Fund} (NO. CCF-ALIMAMA OF 2024009), and the {Natural Science Foundation of Wuhan} (NO. {2024050702030136}), Innovation Scientists and Technicians Troop Construction Projects of Henan Province, China (No. 254000510007), National Key Research and Development Program
of China (No. 2022YFB3102900), the {Xiaomi Young Scholar Program}. 



\bibliographystyle{named}
\bibliography{7Reference}
\newpage
\appendix
\section{Algorithm}
\label{sec:Algorithm}
This section elaborates on the specific operational details of the forward and reverse processes in the implementation of \name. The algorithm is presented below using pseudocode notation:

\begin{algorithm}[h]
    \caption{Reverse Inference Process}
    \label{alg:inference}
    \textbf{Input}: 

        ~~Historical interaction sequence $\mathcal{S} = (i_{1}, i_{2}, ..., i_{m})$
        
        ~~Trained model parameters $\theta$ and $\omega$
        
        ~~Number of Euler computation steps $q$
        
        ~~Scaling factor for Timesteps Schedule $s$

    \textbf{Output}: 

        ~~Recommended next item $i_{m+1}$

    \begin{algorithmic}[1] 
        \STATE $[e_1, e_2, ..., e_m]  = Embedding(\mathcal{S})$
        \STATE Initialize noise $x_n \sim \mathcal{N}(0, I)$
        \STATE Set timestep increment $\Delta t = \frac{1}{q}$
        \STATE $\bm{z}_0 = \bm{x}_n$
        \FOR{$t = 0$ to $1$}
            
            \STATE Predict $f_{\Theta}(\bm{z}_{t}, t)$
            \STATE Compute the vector field $\bm{u}_t(\bm{z}_t | \bm{x}_n) = f_{\Theta}(\bm{z}_{t}, t) - x_n$
            \STATE Update the noise: $\bm{z}_{t +\Delta t}  = \bm{z}_{t} + \Delta t \cdot \bm{u}_t(\bm{z}_t | \bm{x}_n)$
            \STATE $t = t + \Delta t$
        \ENDFOR
        \STATE Compute recommendation scores using Softmax 
        
        $\hat{y} = \text{Softmax}(\bm{z}_1 \cdot Embeddings)$
        \STATE Select the item with the highest score as the recommendation $i_{m+1} = \arg\max_{i \in \mathcal{I}} \hat{y}_i$
        \STATE \textbf{return} $i_{m+1}$
    \end{algorithmic}
\end{algorithm}

\begin{algorithm}[tb]
    \caption{Forward Training Process}
    \label{alg:training}
    \textbf{Input}: 
    
        ~~User interaction sequences $\mathcal{S} = (i_{1}, i_{2}, ..., i_{m})$

        ~~Loss function weight $\alpha$, $\beta$
        
        ~~Scaling factor for Timesteps Schedule $s$
        
        ~~Distribution Hyperparameter $\delta$

    \textbf{Parameter}:
                
        ~~Transformer parameters $\theta$

        ~~Decoder parameters $\omega$
        
    \textbf{Output}: 

        ~~Updated model parameters $\theta$ and $\omega$

    \begin{algorithmic}[1] 
        \FOR{each training epoch}
            \FOR{each batch $\mathcal{B} \subset \mathcal{U}$ in the data}
                \FOR{each user $u \in \mathcal{B}$}
                    \STATE $[\bm{e}_1, \bm{e}_2,\ldots, \bm{e}_m, \bm{e}_{m+1}]  = \texttt{Embedding}(\mathcal{S})$
                    \STATE Sample a timestep $t$ by Mode Sampling
                    \STATE Add noise to the target $\bm{x}_c = \bm{e}_{m+1}$ to obtain $\bm{z}_t$:
                    \STATE \quad $\bm{z}_t = (1-t) \bm{x}_c + t \bm{x}_n$, where $\bm{x}_n \sim \mathcal{N}(0, I)$
                    \STATE Integrating historical interaction sequences:
                    \STATE \quad $\bm{E}_i(\bm{z}_t, t) = \bm{e}_i + \lambda_{i} \odot (\bm{z}_t + t)$

                    \STATE By processing $[\bm{E}_0(\bm{z}_t, t), \ldots, \bm{E}_m(\bm{z}_t, t)]$ within the model, we obtain $f_\Theta(\bm{z}_t, t)$ and $\bm{h}_m$ as defined by Equation~(\ref{equ:f_theta}) and~(\ref{equ:h_n}).
                    
                    \STATE Calculate the FM Loss:
                    \STATE $\mathcal{L}_{FM} = \mathbb{E}_{t, p_t(\bm{z}|\bm{x}_c), p(\bm{x}_c)}||f_\Theta(\bm{z}_t, t) -~\bm{x}_c||_2^2$
                    \STATE Calculate the CE loss $\mathcal{L}_{CE}$ by Equation~(\ref{equ:loss_ce})
                    \STATE Calculate the MSE Loss:
                    \STATE \quad $\bm{\hat{d}} = \text{Decoder}_{\omega}(\bm{h}_m)$
                    \STATE \quad $\mathcal{L}_{MSE} = \| \bm{\hat{d}} - \bm{r} \|^2$
                    \STATE $\mathcal{L} = \mathcal{L}_{FM} +  \alpha  \mathcal{L}_{CE} + \beta \mathcal{L}_{MSE}$ // Total loss
                \ENDFOR
            \ENDFOR
        \ENDFOR
    \end{algorithmic}
\end{algorithm}

\section{Extended Experimental Results}
\label{sec:experienment_appendix}
This section will detail additional experiments conducted beyond the scope of the main text. Specifically, we examine the effects of various Timestep Schedules, analyze the impact of the hyperparameter $s$ in Mode Sampling. We further present results from experiments on the Yelp and Steam datasets, which are not included in the core experimental analysis.

\subsection{Influence of Timesteps Schedule}
\begin{figure}[htbp]
    \centering
    \includegraphics[width=0.48 \textwidth]{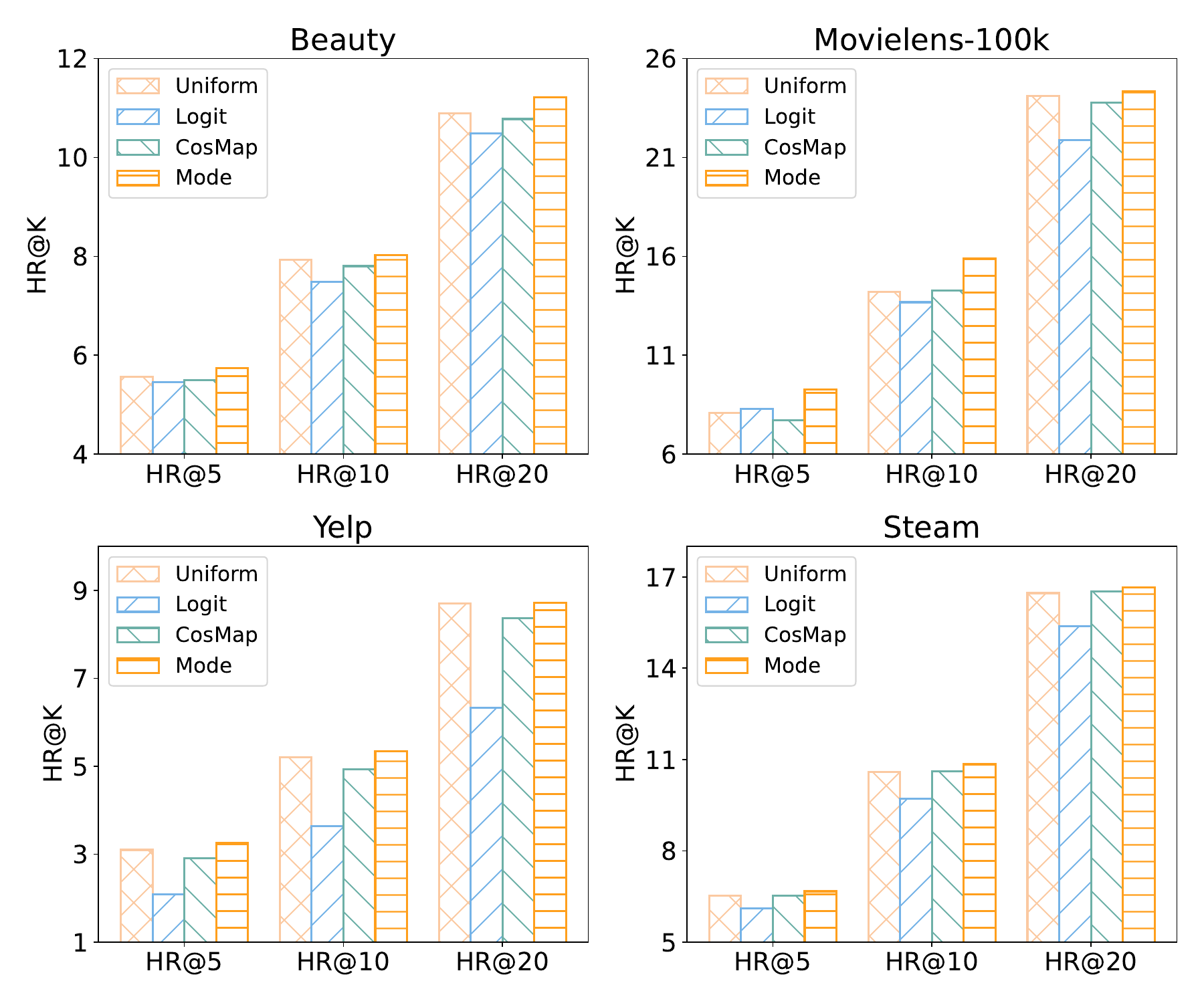}
    \caption{Performance comparison based on different timesteps schedule, measured as percentages(\%)}
    \label{fig:timesteps schedule}
\end{figure}

\name employs a mode sampling technique distinct from traditional Flow Matching, which strategically increases the sampling probability of intermediate steps through an adjustable parameter~$s$. We conducted experimental comparisons of Mode Sampling which we apply (denoted as ``Mode'') against Uniform sampling(denoted as ``Uniform''), Logit-Normal sampling (denoted as ``Logit''), and CosMap sampling(denoted as ``CosMap''), evaluated using the HR@$K$ metrics.
As shown in Figure~\ref{fig:timesteps schedule}, the experimental results indicate that, compared to Uniform Sampling, the noise data from intermediate steps plays a crucial role in the generation of complex data from pure noise. These steps significantly impact the quality of generated samples and should therefore receive more attention.
Conversely, the Logit-Normal and CosMap approach both place excessive focus on the intermediate steps. This leads to a lack of training samples for the initial and final steps, substantially hindering the model's performance.

\subsection{Scaling Parameter within the Timestep Schedule~$s$}
\begin{figure}[h]
    \centering
    \includegraphics[width=0.48 \textwidth]{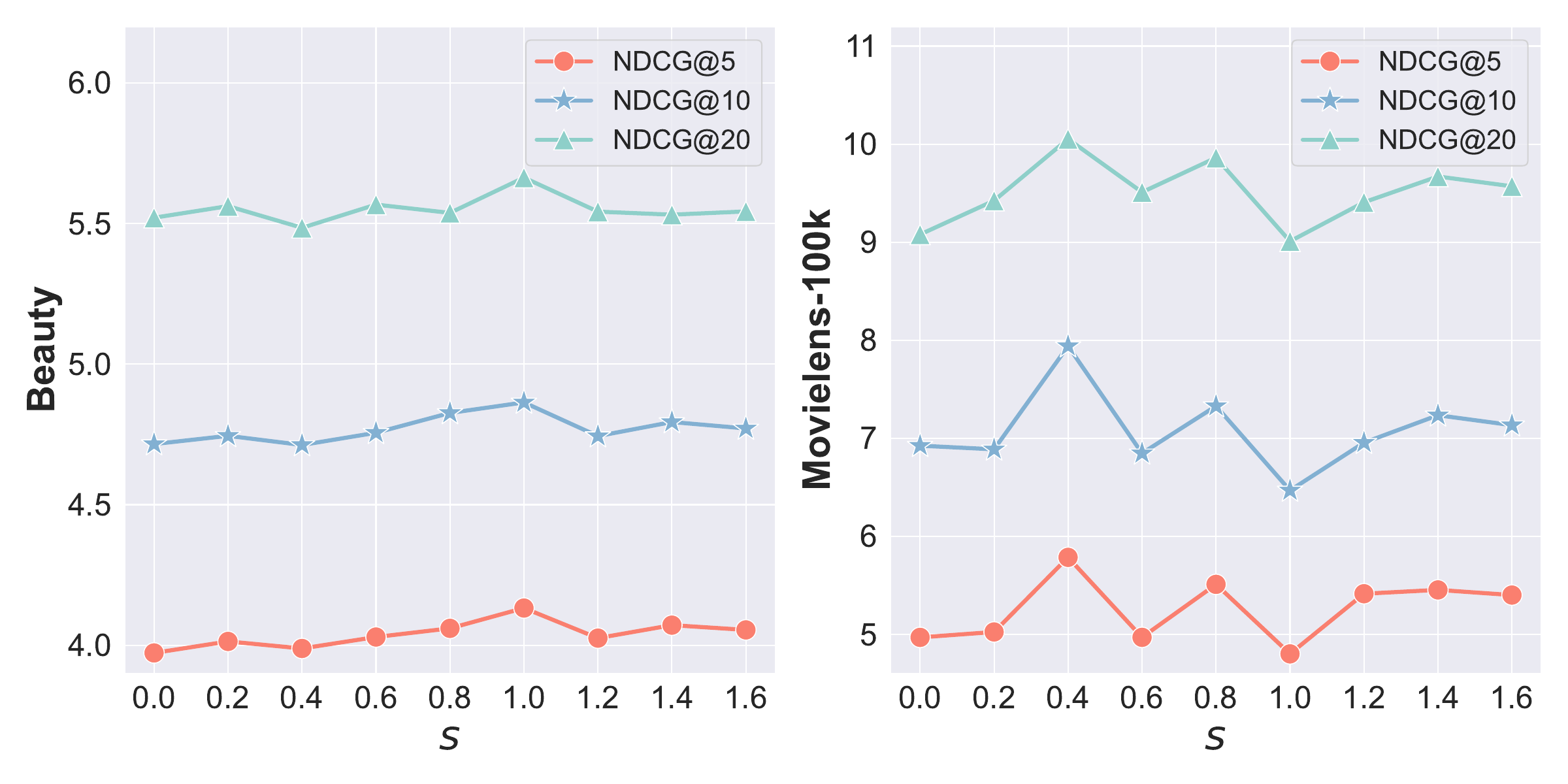}
    \caption{Comparison of model performance across various $s$ on the \textit{Beauty} and \textit{Movielens-100k} dataset, measured as percentages(\%).}
    \label{fig:parameters}
\end{figure}

As mentioned above, Mode Sampling adjusts the sampling probability of intermediate steps through the tunable parameter~$s$. A larger value of~$s$ indicates a higher probability of sampling intermediate steps, while a smaller value brings the sampling process closer to uniform sampling.
We conducted experiments with varying values of~$s$, ranging from 0 to 1.6, to analyze the impact of this parameter on model performance, evaluated using the NDCG@$K$ metrics.
As illustrated in Figure~\ref{fig:parameters}, our experimental results demonstrate that a value of~$s = 0.4$ yields the optimal performance. This suggests that during training, a suitable focus on intermediate steps is necessary while maintaining sufficient attention on the initial and final steps. The balanced increase of sampling probability for intermediate steps effectively enhances model performance.

\subsection{Analytical Experiments on Additional Datasets}
The remaining results of the experiments in Section ~\ref{subsec: analysis experiment}, which were obtained using additional datasets, will be presented in this section.

\paragraph{Influence of Flow Matching Loss}
As shown in Table~\ref{tab:prediction_appendix}, the naive Flow Matching Loss exhibits poor performance on the Yelp and Steam datasets. This suggests that \textbf{directly predicting the vector field} significantly deviates from the core objectives of sequential recommendation, which significantly undermines the model's performance. This further validates the effectiveness of our modified loss for sequential recommendation tasks.

\begin{table}[h]
    \centering
    \resizebox{0.48\textwidth}{!}{%
        \begin{tabular}{lrrrrrr}
            \toprule
            \multicolumn{7}{c}{\textit{Yelp}}  \\ \hline\hline 
             & \textbf{HR@5} & \textbf{HR@10} & \textbf{HR@20} & \textbf{NDCG@5} & \textbf{NDCG@10} & \textbf{NDCG@20} \\
            \hline 
            $v$-prediction & 0.1542 & 0.8532 & 1.0037 & 0.0856 & 0.1579 & 0.5879 \\
            \name  & \textbf{3.3084} & \textbf{5.4421} & \textbf{9.0631} & \textbf{2.1174} & \textbf{2.7855} & \textbf{3.6180} \\
            \toprule 
            \multicolumn{7}{c}{\textit{Steam}}  \\ \hline\hline
             & \textbf{HR@5} & \textbf{HR@10} & \textbf{HR@20} & \textbf{NDCG@5} & \textbf{NDCG@10} & \textbf{NDCG@20} \\
            \midrule
            $v$-prediction & 0.3490 & 0.8531 & 1.4234 & 0.2794 & 0.6941 & 0.9774 \\
            \name & \textbf{6.5254} & \textbf{10.5908} & \textbf{16.4469} & \textbf{4.1878} & \textbf{5.4925} & \textbf{6.9689} \\
            \bottomrule
        \end{tabular}
    }
    \caption{Performance comparison using different Flow Matching losses, measured as percentages(\%). The $v$-prediction approach employs the naive training loss. \name utilizes the modified Flow Matching loss. We highlight the highest-performing metric values in \textbf{bold}.}
    \label{tab:prediction_appendix}
\end{table}

\begin{figure}[h]
    \centering
    \includegraphics[width=0.48 \textwidth]{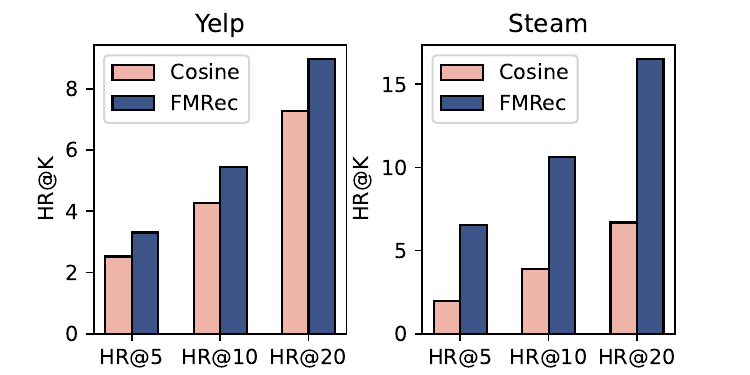}
    \caption{Performance comparison based on different flow trajectories, measured as percentages(\%): ``Cosine'' represents the results obtained using the Cosine trajectory, while ``\name'' denotes the use of the straight trajectory.}
    \label{fig:trajectory_appendix}
\end{figure}

\paragraph{Effectiveness of Straight Trajectories}
As shown in Figure~\ref{fig:trajectory_appendix}, the straight trajectory approach on both the Yelp and Steam datasets also demonstrates enhanced performance, leading to more accurate recommendations when compared with the curved trajectory approach.

\section{Theoretical Benefit of Straight Trajectories}
\label{sec:Theoretical_appendix}
We theoretically demonstrate that straight trajectories inherently avoid discretization errors induced by the Euler method during the reverse process. 
For a step size $\Delta t$ (inverse to the number of Euler steps), the Euler update  
\begin{equation}
\label{equ:Euler}
\small
\bm{z}_{t+\Delta t} = \bm{z}_t + \Delta t \cdot \left.\frac{d\bm{z}}{dt}\right|_{t},
\end{equation}
approximates the true Taylor expansion  
\begin{equation}
\small
\label{equ:Taylor}
\bm{z}_{t+\Delta t} = \bm{z}_t + \Delta t \cdot \left.\frac{d\bm{z}}{dt}\right|_{t} + \frac{(\Delta t)^{2}}{2} \cdot \left.\frac{d^{2}\bm{z}}{dt^{2}}\right|_{t} + \mathcal{O}((\Delta t)^{3}).
\end{equation} 

For \textbf{straight trajectories} (linear function), all higher-order terms ($\frac{(\Delta t)^2}{2} \cdot \frac{d^2\bm{z}}{dt^2}$, etc.) vanish, \textbf{rendering the Euler update exact}. Conversely, \textbf{curved trajectories} incur a truncation error dominated by $\frac{(\Delta t)^2}{2} \cdot \frac{d^2\bm{z}}{dt^2}$, which \textbf{grows quadratically with $\Delta t$}. Thus, with a finite number of steps (i.e., non-infinitesimal $\Delta t$), straight trajectories eliminate this error source entirely, offering a critical advantage in the practical inference process.
\end{document}